\documentclass[aps,nofootinbib]{revtex4}

\usepackage{amsfonts,color}
\usepackage{graphicx,calc,epsfig}

\makeatletter
\@addtoreset{equation}{section}
\makeatother

\newcommand{\ts}{\tau_{\va \! \star}}

\newcommand{\journal}[4]{{\em #1~}#2\,(#3)\,#4}

\newcommand{\pr}{\journal {Phys. Rev.}}

\newcommand{\jmp}{\journal {J. Math. Phys.}}

\newcommand{\cmp}{\journal {Commun. Math. Phys.}}

\newcommand{\cqg}{\journal {Class. Quantum Grav.}}

\newcommand{\np}{\journal {Nucl. Phys.}}

\newcommand{\prep}{\journal {Phys. Rep.}}

\newcommand{\taus}{{\tau_{\va \! \star}}}
\newcommand{\RB}{{\R_{\rm B}}}

\newcommand{\vev}[1]{\left\langle {#1}\right\rangle}

\definecolor{blue}{rgb}{0,0,1}
\definecolor{green}{rgb}{0,1,0}
\definecolor{red}{rgb}{1,0,0}
\definecolor{van}{rgb}{1,0,1}
\definecolor{al}{rgb}{1,1,0}

\newcounter{mnotecount}[section]

\newcommand{\Z}{\mathbb{Z}}

\newcommand{\be}{\nopagebreak[3]\begin{equation}}
\newcommand{\ee}{\end{equation}}
\newcommand{\ba}{\nopagebreak[3]\begin{eqnarray}}
\newcommand{\ea}{\end{eqnarray}}
\DeclareFontFamily{U}{rsfs}{}         
\DeclareFontShape{U}{rsfs}{m}{n}{<5> rsfs5 <6><7> rsfs7          %
  <8><9><10><10.95><12><14.4><17.28><20.74><24.88> rsfs10}{}     %
\DeclareMathAlphabet{\mathfs}{U}{rsfs}{m}{n}                     %
\newcommand{\mfs}[1]{\mathfs {#1}}                               %
\newcommand{\va}{\scriptscriptstyle}
\newcommand{\vani}{\scriptstyle}

\newcommand{\sC}{{\mfs C}}

\newcommand{\sG}{{\mfs G}}

\newcommand{\sH}{{\mfs H}}
\newcommand{\sL}{{\mfs L}}

\newcommand{\sM}{{\mfs M}}

\newcommand{\sX}{{\mfs X}}
\newcommand{\sPP}{{\mfs P}}
\newcommand{\Hp}{{\sH}_{phys}}

\newcommand{\Ha}{{\sH}_{aux}}

\newcommand{\g}{\mathfrak{g}}

\usepackage{psfrag}
\usepackage{amsfonts}
\usepackage{amsmath}
\usepackage{graphics,graphicx,color,calc,epsfig}
\usepackage{euscript}
\newcommand{\beq}{\begin{equation}}
\newcommand{\eeq}{\end{equation}}
\newcommand{\beqa}{\begin{eqnarray}}
\newcommand{\eeqa}{\end{eqnarray}}

\newcommand{\R}{\mathbb{R}}

\newcommand{\om}{\omega}

\begin{document}

\title{Quantization of the   Jackiw-Teitelboim model}

\author{Clisthenis  P. Constantinidis\footnote{ Work supported
   in part by the Conselho Nacional
   de Desenvolvimento Cient\'{\i}fico e
   Tecnol\'{o}gico -- CNPq (Brazil).}$^{,}$\footnote{ Work supported
   in part by the PRONEX project No. 35885149/2006 from FAPES -- CNPq (Brazil).}}
\affiliation{Universidade Federal do Espirito Santo,   Vit\'oria, Brazil}

\author{Alejandro Perez\footnote{Partially supported by the
Programa de Professor Visitante
Estrangeiro, CAPES (Brazil).}}
\affiliation{Centre de Physique Th\'eorique\footnote{Unit\'e Mixte
de Recherche (UMR 6207) du CNRS et des Universit\'es Aix-Marseille
I, Aix-Marseille II, et du Sud Toulon-Var; laboratoire afili\'e
\`a la FRUMAM (FR 2291)}, Campus de Luminy, 13288 Marseille,
France.}

\author{Olivier Piguet$^{*,\dagger}$}
\affiliation{Universidade Federal do Espirito Santo,   Vit\'oria, Brazil}

\date{\today \vbox{\vskip 2em}}

\begin{abstract}
We study the phase space structure of the   Jackiw-Teitelboim model
in its connection variables formulation where the gauge group of
the field theory is given by local ${\rm SL}(2,\R)$ (or ${\rm SU}(2)$ for the
Euclidean model), i.e. the  de Sitter group in two dimensions. In
order to make the connection with two dimensional gravity explicit,
a partial gauge fixing of the de Sitter symmetry can be introduced
that reduces it to spacetime diffeomorphisms. This can be done in
different ways.
Having no local physical degrees of freedom, the
reduced phase space of the model is finite dimensional.  The
 simplicity of this gauge field theory allows for studying
different avenues for quantization,  which may use various (partial) gauge
fixings.
 We show that reduction and
quantization are  noncommuting operations: the representation of
basic variables as operators in a Hilbert space depend on the
order chosen for the latter. Moreover, a representation that is
natural in one case may not even be available in the other leading
to inequivalent quantum theories.
\end{abstract}
\maketitle

\section{Introduction}

  The Jackiw-Teitelboim model~\cite{JT,fuk-kam} is one of the simplest but nontrivial
formulations of General Relativity (GR)
 in two-dimensional space-time with cosmological constant $k$.
Its action is given by
\begin{eqnarray}
S_{\rm JT}=\frac{1}{2}\int d^{2}x \sqrt{-g}\psi (R-2k)\,.\label{2}
\end{eqnarray}
It is  invariant under space-time diffeomorphisms
and leads to the Liouville equation 
\begin{eqnarray}
R-2k = 0\,,
\label{1}
\end{eqnarray}

It contains only a finite number of degrees of freedom, namely one  (here we
assume the space time topology $M=S^1\times \R$).
The model may be quantized in
the original variables of Jackiw and Teitelboim in a canonical framework
including the two first class constraints corresponding to space-time
diffeomorphism invariance~\cite{JT,marc-h}.

On the other hand, one may take profit of its equivalence with a
$BF$ theory~\cite{chamseddine,kum-lie-vas1,kum-lie-vas2},
which has a structure similar to the first order
formulation of four-dimensional GR in Ashtekar's variables. Here,
instead of being the four-dimensional Lorentz group, the gauge
group is that of two-dimensional de Sitter or anti-de Sitter
symmetry ${\rm SO}(1,2)$, or ${\rm SO}(3)$ in the Euclidean de Sitter
case -- or better their covering groups, ${\rm SL}(2,\R)\approx {\rm SU}(1,1)$
or ${\rm SU}(2)$, respectively. The fields are a gauge connection 1-form
$\omega$ and a scalar $\phi$ in the adjoint representation. A
quantization in the Euclidean case was presented
in~\cite{livine-perez-rovelli}, using spin network and spinfoam
techniques~\cite{ash-lewan}.

The canonical formulation of the $BF$ theory gives rise to three
first class constraints whose Poisson Bracket algebra reproduces the
three-dimensional Lie algebra of the gauge group. The quantization
may follow various roads, using some complete or partial gauge
fixing~\cite{kum-lie-vas1,kum-lie-vas2,grum-kum-vas,berg-mey,CLMPR,paper-prepa},
or no gauge fixing at
all~\cite{topol-phys-rep,livine-perez-rovelli}.
In~\cite{CLMPR}, a time gauge has been used, which consists in the
vanishing of the connection component $\omega^0{}_x$ -- which is
interpreted as the space component of the zweibein (2-bein) form
$e^0$ -- with the purpose of simulating the time gauge fixing of
four-dimensional gravity leading to the Ashtekar variables
formulation~\cite{ash-lewan}. This partial gauge fixing leads to a
reduction of the number of first class constraints to two,
corresponding to the space-time diffeomorphism invariance -- namely
one constraint generating the space diffeomorphisms and the other
one playing the role of the Hamiltonian constraint.

Despite of the extensive literature studying the JT model, there is,
to our knowledge, no complete treatment of the quantization of the
Lorentzian sector in its first order formulation using loop variables
(for reviews on other methods see~\cite{Strobl:1999wv,martin}). The main
difficulty is technical: the fact that  the gauge group in that
case is noncompact precludes the possibility of using the standard
quantization techniques that are applicable in the Riemannian
case.

\noindent  The purpose of the present paper is twofold:

On the one hand we study in detail the quantization of the model in
the Lorentzian sector. This is achieved through a minimalistic
application of the techniques developed in~\cite{livine-freidel} which
can also be introduced from the point of view of
\cite{Gomberoff:1998ms}. It is by now known~\cite{josh} that the
general case of a gauge theory with noncompact internal gauge
symmetries presents important difficulties that are not completely
resolved by the methods proposed in~\cite{livine-freidel}.
Interestingly, those difficulties vanish in the 2-dimensional case
and the quantization presented here is well defined.

 On the other hand, we propose a new partial
gauge fixing defined by the vanishing of one component of the
scalar field $\phi$, to compare the corresponding quantum theory
with the theories already constructed~\cite{livine-perez-rovelli}
or under construction~\cite{paper-prepa}. The question makes sense
since it is well-known~\cite{polymersAsh,polymersCorichi} that,
even in theories with a finite number of degrees of freedom,  it
may exist inequivalent quantization of the same classical theory
if some assumption of the von Neumann theorem is invalid, such as
for instance the existence of pairs of self-adjoint operators
``$p\,,\ q$'' obeying Heisenberg commutation relations. In a
gauge theory, inequivalent quantizations can also arise from the
possible inequivalence of the two customary procedures of
quantization consisting in either reducing the unconstrained phase
space and then quantizing or quantizing first and then imposing
the constraints at the quantum level (Dirac procedure). This is
particularly important in our case where, even though the number
of physical degrees of freedom is finite, the unreduced phase
space of the system is infinite dimensional. Consequently, in the
second quantization procedure -- the Dirac procedure -- the von
Newman theorem has no bearing and infinitely  many inequivalent
quantizations exist in principle.

\section{  The Jackiw-Teitelboim model in the $BF$ formulation}
The model can be written as a BF theory~\cite{topol-phys-rep} in 2 dimensions.
 The gauge group $G$ is de Sitter or anti-de Sitter in Riemannian or Lorentzian
space-time. The infinitesimal generators are
\[
J_0:= P_0\,,\quad J_1:= P_1\,,\quad J_2:=\Lambda\,,
\]
with commutation relations
\[
[J_0,J_1]=k J_2\,,\quad [J_0,J_2]=- J_2\,,\quad [J_1,J_2]= \sigma J_1\,,
\]
where $k$ is the cosmological constant and
$\sigma$ is the metric signature, equal to 1 in the Riemannian case and to
$-1$ in the Lorentzian case.
A redefinition of the Lie algebra basis allows to reduce its commutation
rules to:
\beq
[J_i,J_j]= f_{ij}{}^k J_k\,,\quad
\mbox{with}\quad f_{01}{}^2 = 1\,,\ f_{12}{}^0 =\sigma\,,f_{20}{}^1 =1\,,
\quad (i,j,\cdots = 0,1,2)\,,
\label{Lie-G}\eeq
which, for $\sigma= -1$ or 1, is the Lie algebra of SO(3) or SO(1,2). We
shall consider in the following their covering groups SU(2) or
{\rm SL}(2,$\R$). The invariant Killing form $\eta_{ij}$ has the form of a
Euclidian or Minkowskian 3-dimensional metric:
\beq
\eta_{ij} := -\frac{\sigma}{2} f_{ik}{}^l f_{jl}{}^k = {\rm
diag}(\sigma,1,1)\,.
\label{Killing}\eeq

The fields are an
${\rm SU}(2)$ or ${\rm SL}(2,\R)$ connection 1-form $\omega^i$ and
a scalar field $\phi^i$ in the
adjoint representation of the gauge group.
For this point
on we shall denote the
internal gauge group with $G$, and we will use $\mathfrak g$ to designate
its Lie algebra.
We will use the explicit ${\rm SL}(2,\R)$ and ${\rm SU}(2)$
when we specialize to the Lorentzian or Riemannian models
respectively.
The components $\omega^0$ and $\omega^1$ are interpreted as the zweibein
components and $\omega^2$ as the rotation, respectively Lorentz
conection.

The theory is the 2d version of BF theory,
and can be seen as the $g\mapsto 0$ limit ($g$ being the coupling
constant) of 2d Yang Mills theory.
Its action takes the form
\be S=\int_{\sM} \phi^i F^{j}(\omega)\eta_{ij}\,,
\ee
and the field equations are
\beq\begin{array}{l}
d_{\omega}\phi^i := d\phi^i +f_{jk}{}^i\omega^j \phi^k = 0\,, \\
F^i := d\omega^i + \frac{1}{2} f_{jk}{}^i\omega^j \omega^k   = 0 \,,
\end{array}\label{bf1}\eeq
where $d_\omega$ is the $\omega$-covariant exterior
differential and $F^i$ the curvature 2-form of the connection.
At first sight the action is invariant under two
kinds of gauge transformations: the conventional Yang-Mills-like
local $G$ transformations generated by a  Lie algebra valued scalar
field $\lambda$
\begin{eqnarray}\nonumber
&&\delta_{\lambda} \omega = d_{\omega}\lambda\,, \\
&&\delta_{\lambda} \phi  = [\phi ,\lambda]\,, \label{infini}
\end{eqnarray}
whose exponentiated version gives
\begin{eqnarray}\nonumber
&&\omega' = a\omega a^{-1}+ad{a^{-1}}\,, \\
&& \phi'  = a\phi a^{-1}\,, \label{fini}
\end{eqnarray}
for $a=\exp(\lambda)\in G$. The active diffeomorphisms  are generated
by a vector field $v$
\begin{eqnarray}
\nonumber &&\delta_{v} \omega = \sL_{v}\omega\,, \\
&&\delta_{v} \phi  = \sL_{v}\phi\,,
\end{eqnarray}
$\sL_{v}$ being the Lie derivative.  However, on shell a
diffeomorphism generated by $v$ is the same transformation as a
local $G$-transformation generated by the field
\begin{equation}
    \lambda^i=v^a \omega_{a}^i\,,
\end{equation}
as can be easily checked by writing these equations in components
and using the equations of motion.  Therefore, in this theory the
diffeomorphisms (acting on the space of solutions) can be
considered as a subgroup of the local $G$ gauge transformations.

Solutions are given by flat connections $\omega=gdg^{-1}$ for
$g(x)\in G$, and (covariantly) constant $\phi$ fields. Locally,
one can choose a gauge so that the connection $\omega=0$. In this
gauge the equation $d_{\omega}\phi=0$ implies that $\phi={\rm
constant}$. This is particularly important for the Lorentzian case
since it implies that the causal type  of $\phi$ (thought as a vector
in three dimensional Minkowsky internal geometry) cannot change in
the classical solutions.  This conclusion is a global one
since no gauge transformation $\phi\rightarrow g\phi g^{-1}$
 can send a time-like $\phi$ into a space-like one or viceversa.
In fact, this property is taken over to the quantum theory where
we will show that  super-selection sectors associated to $\phi$
being space-like or time-like appear.

\section{The Hamiltonian formulation}

When $\sM=S^1\times R$ one can quantize the theory in the
canonical framework. General topologies can in principle be
considered in the path integral approach. The Hamiltonian
formulation is obtained through the standard $1+1$ spacetime
decomposition. More precisely, one introduces an arbitrary
foliation of $\sM$ by choosing a time function. In terms of this
foliation the action becomes
\be S=\int dt \int_{S^1} dx\,( \phi_i
\dot{\omega}^i+\omega_t ^i D\phi_{i})\,.
\ee
 We use the notations
$D\phi^i:=\partial\phi^i+f_{jk}{}^i\om^j\phi^k$,
$\omega^i:=\omega^i_x$, $\partial:=\partial_x$,
$x$ being the space coordinate.
The Poisson bracket among
the phase space variables is
\[
\{\omega^j(x) ,\phi_i(y)\}=\delta^j_i\delta(x-y)\quad\mbox{or}\quad
\{\omega^j(x) ,\phi^i(y)\}=\eta^{ij}\delta(x-y)\,.
\]
We have three first class constraints~\cite{dirac,henneaux-teit,tyutin}
corresponding to the three
components of the Gauss law $g_i:=D\phi^i=0$. Explicitly these
components are:
\ba\nonumber && g_0=\partial \phi^0 + \sigma(\omega^1\phi^2-\omega^2\phi^1)\approx 0\,, \\
\nonumber && g_1=\partial \phi^1+\omega^2\phi^0-\omega^0\phi^2 \approx 0\,,\\
&& g_2=\partial \phi^2+\omega^0\phi^1-\omega^1\phi^0\approx 0\,.
\label{gauss}\ea
The smeared Gauss constraint $g(\alpha)\equiv\int_{S^1} {\rm Tr}[\alpha
D\phi]$ for $\alpha\in \g$ are first class -- they satisfy the
Poisson bracket identity
$\{g(\alpha),g(\beta)\}=g([\alpha,\beta])$ -- and generate
infinitesimal $G$-gauge transformations:
\be
\{g(\alpha),\omega^i\}=\partial\alpha^i +
f_{jk}{}^i\omega^j\alpha^k\,,\quad
\{g(\alpha),\phi^i\} = f_{jk}{}^i\phi^j\alpha^k\,.
\label{G-transf}\ee
 There are therefore three
local first class constraints for the three configuration
variables $\omega_a^i$. Thus the naive counting of degrees of
freedom gives zero physical degrees of freedom. However, the naive
counting is only sensitive to local excitations. The theory has
indeed global degrees of freedom. In particular, if $M=S^1\times
\R$,  an algebraic basis for the gauge invariant (Dirac) observables is given by
\be
O_1=\phi_i\phi^i\,,\ \ \ O_2={\rm Tr}[P \exp(-\int_{S^1} \omega)]\,.\ \ \
\label{class-dirac-obs}\ee
The physical phase space being therefore 2-dimensional, the  theory
has a single (global) degree of freedom.
 We note for further use that the quantity
\be
Q:= \frac{\omega^i\phi_i}{\phi^i\phi_i}
\label{abelian-conn}\ee
transforms as an Abelian connection under the special gauge
transformations which leave $\phi$ invariant:
\be
\{ g(\alpha_{\rm Abel}),Q \} \approx \partial a\,,\quad
\mbox{with}\quad \alpha_{\rm Abel}^i=a\phi^i\,.
\label{abelian-transf}\ee
This holds up to the constraint $\partial(\phi_i\phi^i)\approx 0$ which
follows from (\ref{gauss}).

\subsection{Partial gauge fixings}\label{Partial gauge fixings}

Spacetime diffeomorphisms  are hidden inside the larger gauge
group of $BF$ theory that in two dimensions corresponds to the
local $G$ transformations given in (\ref{infini}). Notice that the
former are generated by the two components of a vector field in
$\sM$ while the latter are generated by the three components of
$\lambda\in\g$. In this section we partially gauge fix the
symmetries of BF theory in order to establish a more direct
relationship with diffeomorphisms, and hence emphasize the
relationship of the model with two dimensional gravity.

 We will partially gauge fix the $G$ gauge symmetry by
requiring the fourth constraint \be g_3=\phi_i n^i\approx
0\label{gaugefix}\,,\ee where $n^i$ is a fixed normalized vector in
the internal space.
 Before going into the technical details of the constraint
algebra let us discuss the geometric interpretation of the partial
gauge fixing introduced by the above equation. Due to the fact
that $\phi\cdot\phi$ is a Dirac observable (a constant of motion)
one can separate the analysis into three distinct (dynamically
independent) cases in the Lorentzian case: $\phi\cdot\phi>0$,
$\phi\cdot\phi=0$ and $\phi\cdot\phi<0$ (in the Riemannian case
 there is only the first sector).
The following discussion is restricted to the Lorentzian sector.
\subsubsection{The  ``space-like'' sector: $\phi\cdot\phi>0$}

In that case the  good choice of gauge fixing corresponds to
$n^i=$time-like. The condition (\ref{gaugefix}) is expected to
reduce the group ${\rm SL}(2,\R)$ to a two dimensional subgroup. A
moment of reflection shows that this is given by the cartesian
product of the ${\rm U}(1)\subset {\rm SL}(2,\R)$ that leaves invariant $n^i$,
 and  the little group (the boosts) leaving invariant the vector
$\phi^i$. All this will become transparent in the following.

The choice $n^i=$space-like leads to a degenerate situation for
phase space points where $\phi^i\propto n^i$, as on these points
the little groups associated to $n^i$ and $\phi^i$ coincide. This
leads to complications that we will not analyze in this work.

To simplify notation we can take $n^i=(1,0,0)$ which  simply turns
(\ref{gaugefix}) into simply $g_3=\phi^0\approx 0$. With this
choice,  the matrix
$G_{\alpha\beta}=\{g_{\alpha},g_{\beta}\}$ becomes
\be
G(x,y)=\left[\begin{array}{cccc} 0&0&0&0\\ 0&0&0&-\sigma\phi^2\delta(x-y)\\
0&0&0&\sigma\phi^1\delta(x-y)\\
0&\sigma\phi^2\delta(x-y)&-\sigma\phi^1\delta(x-y)&0\end{array}\right]
\ee
In order to isolate the second class part of the previous
constraints we make the following redefinition
\ba\nonumber && C_0=g_0=\partial \phi^0
   +\sigma(\omega^1\phi^2-\omega^2\phi^1)\approx 0 \,,\\
\nonumber && C_1=\phi^1 g_1+\phi^2 g_2=\phi^1(\partial \phi^1
 +\omega^2\phi^0-\omega^0\phi^2)+\phi^2(\partial \phi^2
 +\omega^0\phi^1-\omega^1\phi^0) \approx 0\,,\\
&& \nonumber C_2=\phi^1 g_1-\phi^2 g_2=\phi^1(\partial
\phi^1+\omega^2\phi^0-\omega^0\phi^2)-\phi^2(\partial
\phi^2+\omega^0\phi^1-\omega^1\phi^0) \approx 0\,,\\ &&
C_3=g_3=\phi^0\approx 0 \,.
\ea
With this definition the matrix
$C=\{C_{\alpha},C_{\beta}\}$ is block diagonal, namely,  up to
terms involving constraints:
\be
C \approx \left[\begin{array}{cccc} 0&0&0&0 \\ 0&0&0&0\\
0&0&0&-2\sigma\phi^2\phi^1 \delta(x-y)\\
0&0&2\phi^2\sigma\phi^1\delta(x-y)&0\end{array}\right]
\ee
This
implies that the pair $(C_2,C_3)$ is second class and can
therefore explicitly solved. For instance  they can be used to
solve for $\omega^0$. Namely
\be
\phi^0=0\,,\quad
\omega^0=\frac{1}{2}\left(\frac{\partial
\phi^1}{\phi^2}-\frac{\partial \phi^2}{\phi^1}\right)\,.
\ee
 The first class ones become:
\ba
&&\nonumber
C_0=\omega^1\phi^2-\omega^2\phi^1=\epsilon_{AB}\omega^A\phi^B\approx
0 \,,\\ && C_1=\frac{1}{2}\partial (\phi^A\phi^A) \approx 0\,,
\label{C-first-class}\ea
where
$A,B=1,2$. It is easy to see that the Dirac bracket among the
remaining variables is
\be
\{\omega^{B}(x),\phi^A(x)\}_{\va D}=\delta^{BA}\delta(x-y)\,.
\label{Dirac-bracket}\ee
Direct computation  shows that the algebra  of the
first class constraints is Abelian.
More precisely, if we define
\be
C_0(a)=\int_{S^1} dx a(x) C_0(x)\,,\quad \mbox{and}\quad
C_1(b)=\int_{S^1} dx b(x)C_1(x)\,,
\label{C(a)}\ee
we have the following Dirac bracket algebra:
\be
\{C_0(a),C_0({b})\}_{\va D}=0=\{C_1(a),C_1({b})\}_{\va D}
\ee
and
\be
\{C_0(a),C_1({b})\}_{\va D}=-\frac{1}{2}\int_{S^1} dx\int_{S^1} dy \
a(x)\partial b(y) \ \epsilon_{AB} \phi^A(x) \{\omega^{B}(x),
\phi^C \phi^C(y) \}_{\va D}=0\,.
\ee
The gauge transformations generated by
the first class constraints are
\ba
\nonumber && \delta_{\va
(0)}\omega^A=\{\omega^A,C_0(a)\}_{\va D}=-\sigma a\epsilon^{BA}\omega^B\,,\\
&&\delta_{\va
(0)}\phi^A=\{\phi^A,C_0(a)\}_{\va D}=-\sigma a\epsilon^{BA}\phi^B\,,
\ea
which
correspond to  local U(1) internal rotations, and
\ba
\nonumber &&
\delta_{\va
(1)}\omega^A=\{\omega^A,C_1(a)\}_{\va D}=-\phi^A \partial a\,, \\
&&\delta_{\va (1)}\phi^A=\{\phi^A,C_1(a)\}_{{}_{\va D}}=0\,.
\ea
What is
the geometric meaning of the transformation generated by $C_1$?
Recall that $C_1\equiv \phi^1g_1+\phi^2 g_2$ which is nothing else
than the expression in the gauge (\ref{gaugefix}) of $\phi^i g_i$.
This is precisely the generator of the internal ``boosts'' leaving the
``space-like'' field
$\phi^i$ invariant, as anticipated above.

The Riemannian theory is fully described by the equations of this
section. The only change is that both $C_0$ and $C_1$ generate
${\rm U}(1)$ transformations in that case.

\subsubsection{The ``time-like'' sector: $\phi\cdot\phi<0$}\label{time-like sector}

 The gauge fixing analogous to the previous case
would now be defined with $n^i$ spacelike and we
would
expect condition (\ref{gaugefix}) to reduce the group ${\rm SL}(2,\R)$
to the cartesian product of the boosts leaving invariant $n^i$, and of
the little group (the ${\rm U}(1)$ rotations) leaving invariant the
vector $\phi^i$. It seems however difficult to control the positivity of
$\phi\cdot\phi$, which is no more automatic.
A more appropriate choice is to take $\phi^i(x)=\phi(x)u^i$
where $u$ is some fixed time-like vector.
With the choice $u=(1,0,0)$, this amounts to add to the constraints (\ref{gauss}) -- taken with $\sigma=-1$ -- the new constraints
\[
g_3=\phi^1 \approx 0\,,\quad g_4=\phi^2 \approx 0\,.
\]
The $5\times5$ Poisson bracket matrix
$G_{\alpha\beta}=\{g_{\alpha},g_{\beta}\}$, $\alpha,\beta=0,\cdots,4$
 reads (up to constraints):
\be
G(x,y) \approx \left[\begin{array}{ccccc}
0&0&0&0&0\\
0&0&0&0&1\\
0&0&0&-1&0\\
0&0&1&0&0\\
0&-1&0&0&0\end{array}\right]\phi(x)\delta(x-y)\,.
\ee
Only $g_0$ is first class, the remainder four constraints being of
second class. Eliminating them through the Dirac procedure, we are left
with the strong conditions
\[
\phi^A=0\,,\quad\omega^A=0\,,\quad A=1,2\,,
\]
and the Dirac bracket algebra (with $\omega:=\omega^0$)
\be
\{\om(x),\phi(y)\}_{\va D}=-\delta(x-y)\,.
\label{temp-case-Dirac-bracket}\ee
The first class constraint reads
\be
g_0(x)=\partial\phi\,\quad
\mbox{or, in integral form:}\quad g_0(a) = -\int_{S^1}dx\,\partial a\,\phi\,.
\label{g-0-constraint}\ee
and generates the U(1) gauge rotations which leaves $\phi^i$ invariant:
\[
\{g_0(a),\phi(y)\}_{\va D} = 0\,,\quad
\{g_0(a),\om(x)\}_{\va D}=\partial a\,.
\]
$\omega$ plays the role of the U(1) connection. We note that the
connection $Q$ (\ref{abelian-conn}) is equivalent to $\omega$, since
$Q=\omega/\phi$ and $\phi$ is constrained to be a constant.


\subsubsection{The ``null'' sector: $\phi\cdot\phi=0$}

We add to the constraints (\ref{gauss})
 -- taken with $\sigma=-1$ -- the partial gauge fixing constraint
\[
g_3=\phi^0-1 \approx 0\, .
\]
plus an additional constraint $g_4=\phi\cdot\phi\approx 0$
which imposes the null condition. Notice that $g_4$ commutes with
the Gauss constraints and with $g_3$ so it is automatically
first class. The constraint algebra is the same as in the spacelike sector.
The first class constraints are again $C_0=\epsilon_{AB} \phi^A\omega^B$ and
$C_1=\partial (\phi^A\phi_A)/2
 - \epsilon_{AB} \phi^A\omega^B$ and $g_4=\phi^A\phi_A-1$
The  constraint system is clearly reducible as $C_1$ is a combination of the
other. Following the standard procedure we drop the constraint $C_1$.
Classical solutions are maps from $S^1$ to $S^1$, due to the action of
$C_0$ only the homotopy class of these maps is physically meaningful.
The quantization of the null sector is outside the scope of this paper
\footnote{For a study of the topological properties of the
classical solutions of a more general type of models of which this case is a
particualr one see \cite{martin}}.

\subsubsection{Diffeomorphisms, Virasoro  and  Abelian generators}

 For simplicity the following analysis is performed in the
$\phi\cdot\phi>0$ sector. Let us define
\be
\Gamma=\frac{\epsilon_{AB}\phi^B\partial\phi^A}{\phi^C\phi^C}\,,
\ee
solution of the equation
\be
\partial\phi^A - \epsilon^{AB}\Gamma \phi^B=0\,.
\label{def-Gamma},\ee
 an analog to the torsion free connection of general relativity in the first order
formulation.
We may check that $\Gamma\approx\omega_0$ and that (\ref{def-Gamma}),
with $\Gamma$ replaced by $\omega_0$,
is a constraint, a combination of the
first class constraint $C_1$ and of the second class one $C_2$.

 One can introduce variables invariant under the gauge group
generated by $C_0$ as follows:
\be \label{u1variables}
\Pi\equiv\frac{1}{2}\phi^A\phi^A \ \ \ {\rm and} \ \ \
Q\equiv\frac{\phi^A\omega^A}{\phi^C\phi^C}\,,
\ee
 obeying the following equation:
\ba\label{lema}
\int_{S^1} dx a(x)
\{\Gamma(x),Q (y)\}_{\va D}=0\,,
\ea
for arbitrary $a(x)\in C^1(S^1)$.
$\Pi$ corresponds, in the $\phi^0=0$ gauge, to the invariant $O_1$
defined in (\ref{class-dirac-obs}),
 whereas $Q$ is the Abelian connection (\ref{abelian-conn}).
The meaning of these quantities will become clearer in the next section.

The constraints $C_0$ and $C_1$ are scalar densities of weight one.
This is why they are naturally smeared with scalar functions $a$
and $b$ in order to produce  coordinate independent quantities
$C_0(a)$ and $C_1(b)$ respectively. We would like now to define an
equivalent set of constraints that would be suitably smeared with
vector fields of $S^1$. In order to do this one needs first to
define vector density constraints which can be achieved by
multiplying the original ones by density one phase space
functions. Without further motivation we  redefine the constraints
as:
\ba
&&\nonumber V_1=-\Gamma C_0=
-\frac{\epsilon_{AB}\phi^B\partial\phi^A}{\phi^C\phi^C}
\epsilon_{DE}\omega^{D}\phi^{E}\approx 0\,,\\
&&   V_2= - Q C_1=  -
\frac{1}{2}\frac{\phi^A\omega^A}{\phi^C\phi^C}\partial(\phi^B\phi^B)\approx
0\,.
\ea
These are vector densities of weight one (or scalar densities
of weight two). As long as we are away from configurations for
which $\Gamma=0$ or $Q=0$ the previous constraints define the same
constraint surface.  Assuming we have two vector fields $\alpha$ and
$\beta$, we define the smeared versions of the previous constraints
in the obvious manner. Then one has that
\be
\{V_1(\alpha),V_2(\beta)\}_{\va D}=0\ee and
\ba &&\nonumber \{V_1(\alpha),V_1(\beta)\}_{\va D}=V_1([\alpha,\beta])\,, \\
&&\{V_2(\alpha),V_2(\beta)\}_{\va D}=V_2([\alpha,\beta])\,.
\ea
where
$[\alpha,\beta]$ is the  vector field commutator. Therefore
 $V_1$ and $V_2$  commute
with respect to each other, each of them satisfying a classical Virasoro algebra.
They look like diffeomorphism generators, however none of the
two generates diffeomorphisms of $S^1$.
The combination that does
this is: \be D=V_1+V_2\,.\ee  The analog of the Hamiltonian
constraint of gravity  then is \be H=V_1-V_2\,.\ee These satisfy the
`gravity' algebra \ba && \nonumber
\{D(\alpha),D(\beta)\}_{\va D}=D([\alpha,\beta])\,,\\
&& \nonumber \{H(\alpha),H(\beta)\}_{\va D}=D([\alpha,\beta])\,,\\
 && \{H(\alpha),D(\beta)\}_{\va D}=H([\alpha,\beta])\,.\ea
The constraint $D(\alpha)$ generate standard diffeomorphisms as
can be checked by a direct calculation, namely \ba &&
 \{D(\alpha), \phi^A \}_{\va D}= \alpha \Gamma
\epsilon^{AB}\phi^B   \approx - \alpha \partial \phi^A  =
 - {\sL}_{\alpha} \phi^A \label{sdifffi}\,,\\
&&
 \{D(\alpha), \omega^A \}_{\va D}= \alpha \Gamma \epsilon^{AB}\omega_B
  - \partial (\alpha Q) \phi^A
 \approx - ( \alpha \partial \omega^A   + \partial \alpha  \omega^A  )
= - {\sL}_{\alpha}
 \omega^A\,,
\label{sdiffomega} \ea where the weak equalities in
(\ref{sdifffi}) and (\ref{sdiffomega}) mean the insertion of the
constraint equations.   Similarly, $H$ generates time evolution up
to space diffeomorphisms and field equations.

\subsection{Full Reduction}

Notice,  from  (\ref{u1variables}), that
$(Q(x),\Pi )$ are a conjugate ${\rm U}(1)$-invariant fields, i.e., they
strongly commute with $C_0$), and
\be\{Q(x),\Pi(y) \}_{\va D}=\delta(x-y)\,.
\ee
Since $\Pi$ also commutes with $C_1$, it represents a strong Dirac
observable of the model. So we introduce \be \sPP\equiv\int_{S^1}
\Pi\,.\ee On the other hand $Q$ does not commute with $C_1$ but it
 transforms as an Abelian connection -- c.f. (\ref{abelian-transf}):
\be
\delta_{\va (1)}
Q=\{Q,C_1(a)\}_{\va D}=-
\partial a\,.
\ee

 Therefore, it is quite easy
to define a strong Dirac observable
\be
\sX\equiv\int_{S^1} Q\ee
so that \be\{\sX,\sPP\}_{\va D}=1\,.
\label{XP}\ee
The connection $Q$ is an $\R$-connection (associated to the boost
structure group) in the
$\phi\cdot\phi>0$ sector, while it becomes a ${\rm U}(1)$-connection in
the $\phi\cdot\phi<0$ sector.

\subsection{Time-gauge reduction}\label{Time-gauge reduction}

In this section we describe the results obtained in
\cite{CLMPR} where the `temporal gauge' was considered. This
gauge fixing is based in the one taken in four dimensions, through
which we can obtain a compact gauge group as the residual
symmetry, as described in~\cite{ash-lewan}. It consists
in making the  zweibein component $\chi$ $:=$ $\omega^0_x$ vanish and it is
implemented as an extra constraint $\chi\approx 0 $.  The action
then reads
\be
S=\int dt \int_{S^1} dx\, (\phi_i
\dot{\omega}^i+\omega_t ^i D\phi_{i} + B\,\chi)   \,,
\ee
and  again the Dirac
method is used in order to  eliminate the second class constraints. The
remaining constraints, namely $ \mathcal{G}_0'$ and $\mathcal
{G}_1'$ are
\begin{eqnarray}
\mathcal{G}'_{0}(x)&=&(\omega^{1}_{x})\mathcal{G}_{0}(x) =\sigma
\omega^{1}_{x}\partial_x\!\!\left(\frac{\partial_x\phi^2}{\omega^1_x}\right)
+ k(\omega^1_x )^2\phi^2 - \omega^1_x{\omega^2_x}\phi^1\,, \label{g0}\\[0,2 cm]
\mathcal{G}'_{1}(x)&=& \omega^{1}_{x}\mathcal{G}_{1}(x) =
\omega^{1}_{x}\partial_{x}\phi^{1}+{\omega^2_x}\partial_{x}\phi^2\,.
\label{g1}
\end{eqnarray}
The Dirac bracket algebra of these contraints is closed:
\begin{eqnarray}
\left\{\mathcal{G}'_{0}(\epsilon),\mathcal{G}'_{0}(\eta)\right\}_{\va D}\!
&=&\sigma\,\mathcal{G}'_{1}([\epsilon,\eta])\,,\nonumber\\
\left\{\mathcal{G}'_{0}(\epsilon),\mathcal{G}'_{1}(\eta)\right\}_{\va D}\!
&=&-\,\mathcal{G}'_{0}([\epsilon,\eta])\,,
\label{alg3}\\
\left\{\mathcal{G}'_{1}(\epsilon),\mathcal{G}'_{1}(\eta)\right\}_{\va D}\!
&=&-\mathcal{G}'_{1}([\epsilon,\eta])\,, \nonumber
\end{eqnarray}
where $[\epsilon,\eta]\,=
(\epsilon\partial_{x}\eta-\eta\partial_{x}\epsilon)\,$, which
confirms that $\mathcal{G}'_{0}$ and $\mathcal{G}'_{1}$ are first
class. In fact, time diffeomorphisms are generated by
$\mathcal{G}'_{0}$ up to  constraints, up to field equations
(``on-shell realization''), and up to a compensating local Lorentz
transformation which takes care of the time gauge condition. The
second unbroken invariance is that of space diffeomorphisms,
generated by $\mathcal G'_1$.

Observe also that $\mathcal{G}'_{0}$ and $\mathcal{G}'_{1}$ are
scalar densities of weight 1 and this ensures that they form a
closed Lie algebra (\ref{alg3}) -- in contrast with gravity in
higher dimensions where the algebra closes with field dependent
structure
``constants''~\cite{rovelli_book,ash-lewan,thiemann_book}. Such a
feature is characteristic of 2-dimensional theories with general
covariance, such as the bosonic string in the approach
of~\cite{thiemann-string}.

A new redefinition
\begin{eqnarray}
\mathcal{\sC_+}&=&\frac{\sqrt{-\sigma}}{2}\,\mathcal{G}'_{0}-\frac{1}{2}\,
\mathcal{G}'_{1}\,,\\
\mathcal{\sC_-}&=&-\frac{\sqrt{-\sigma}}{2}\,\mathcal{G}'_{0}-\frac{1}{2}\,
\mathcal{G}'_{1}\,,
\end{eqnarray}
leads to the algebra
\begin{eqnarray}
\left\{{\sC}_+(\epsilon),{\sC}_+(\eta)\right\}_{\va D} \!&=
&{\sC}_+([\epsilon,\eta])\,,\nonumber\\
\left\{{\sC}_-(\epsilon),{\sC}_-(\eta)\right\}_{\va D}
\!&=&{\sC}_-([\epsilon,\eta])\,,\\
\left\{{\sC}_+(\epsilon),{\sC}_-(\eta)\right\}_{\va D}\!&=&0\,.
\nonumber 
\end{eqnarray}
which shows a factorization in two classical Virasoro algebras,
 as in the $\phi^0=0$ gauge of Subsection \ref{Partial gauge
fixings}.

\section{Quantization}

Quantization prescriptions take a classical theory as an input and
are supposed to give us a quantum theory. As it is well known (and
should be expected) this recipe is not complete and leads often to
inequivalent quantum theories. Consequently, it is instructive to
have explicit examples available to illustrate this point. Here we
show that different natural choices lead to inequivalent quantum
theories in our ultra simple gauge field theory.

\subsection{Dirac quantization}\label{Dirac_quant}

The physical Hilbert space is given by gauge invariant square
integrable functions of $G$. They are therefore class functions
$f(g)=f(aga^{-1})$ for all $g,a\in G$. There are important
differences between the quantization of the Riemannian and
Lorentzian cases. Therefore, we shall treat each case separately
in this section.

\subsubsection{The Riemannian case, $G={\rm SU}(2)$}

 This case is treated in great detail in~\cite{livine-perez-rovelli,isler}.
We briefly review the results
here. As in the case of LQG one shifts emphasis from smooth
connections to holonomies
\[
g_{\gamma}[\omega]={\rm P}\exp(-\int_\gamma \omega)
\]
along oriented paths $\gamma\subset \Sigma$. In the context of the
Dirac quantization program one first introduces an auxiliary
Hilbert $\Ha$ space where the holonomy $g_{\gamma}[\omega]$  and
the scalar fields $\phi^i$ are represented  as operators. In the
present case,  once holonomies and $\phi^i$ have been chosen as
fundamental variables, there is a natural choice of $\Ha$ where
diffeomorphisms are unitarily implemented. In dimension higher than
two this representation is unique (up to unitary equivalence); however, for
technical reasons the uniqueness theorem \cite{hanno}, does not apply to
the two dimensional case: uniqueness remains an open question.

This
Hilbert space is given by the Cauchy completion in a appropriate
topology of the  algebra of functionals of the connection that
depend on the holonomy of $\omega$ along paths that are edges of
arbitrary graphs in $\Sigma$ (the details of this construction
turn out not to be important for our simple model). In a second
step one promotes the Gauss constraints (\ref{gauss}) to self
adjoint operators satisfying the appropriate quantum constraint
algebra, and finally looks for  the physical Hilbert space
$\Hp\subset \Ha$ defined by the kernel of the quantum constraints.

Due to the simplicity of our model these steps can be shortcut and
we can directly construct the physical Hilbert space $\Hp$ in one
stroke. The logic is as follows: The Gauss constraints generate
infinitesimal ${\rm SU}(2)$  gauge transformations. Elements of $\Hp$ are
gauge invariant functions of the Holonomy. As in LQG this
restricts the set of possible graphs on which states are defined
to  closed ones. In our case there is a unique close graph
corresponding to the entire initial value surface $\Sigma$.
Therefore, the physical Hilbert space is given by functions of the
holonomy $g[A] \in {\rm SU}(2)$ around $\Sigma=S^1$ which are in the
kernel of the Gauss constraints (\ref{gauss}). Those are square
integrable functions which are invariant under
$\psi(g)=\psi(aga^{-1})$ for any $a\in {\rm SO}(3)$ (this invariance is
the residual  gauge action on the based point from where the
holonomy around the Universe is defined). The latter are the
so-called class functions $\psi(g)\in \sL^2({\rm SO}(3))/\sG\subset
{\sL^2({\rm SO}(3))}$.  Using the Peter-Weyl theorem they can be written
as \be \Psi(g)=\sum_{j} (2j+1) \psi_j {\rm Tr}(D^j[g]),\ee where
$D^j[g]$ are unitary irreducible representations of ${\rm SU}(2)$ , and
$\psi_j=\int dg \overline{{\rm Tr}(D^j[g])} \Psi[g]$
 with $dg$ the Haar
measure. Consequently there is a natural construction of the
physical Hilbert space based on the following choice of inner
product
\[
\langle\Psi,\Phi\rangle=\int dg \overline {\Psi(g)}\Phi(g)\,.
\]
Any gauge invariant function ($\psi(g)=\psi(aga^{-1})\in\Hp$) can
be expanded in terms of the characters $\chi_j(g)={\rm
Tr}[D^j(g)]$. This are the strict analog of the so-called
spin-network states of LQG. A fact that will become more important
in the Lorentzian case is that class functions can be thought  of
as functions of the (unique up to conjugation) Cartan subgroup
$H\subset {\rm SO}(3)$. More concretely, any ${\rm SU}(2)$  rotation can be
characterized by a ${\rm U}(1)$ rotation by an angle $\nu\in [0,\pi]$
around an  axis defined by a unit vector $\hat n\in R^3$. The
latter is entirely defined by a point on the unit  2-sphere labelled
by spherical coordinates $\theta\in [0,\pi]$ and $\phi\in
[0,2\pi]$. One can express the ${\rm SU}(2)$  Haar measure in these
coordinates as follows:
\[
dg = \frac{1}{2\pi^2} \sin^2(\nu) (\sin(\theta) d\theta d\phi) d\nu\,.
\]
Now any class function depends only on the coordinate $\nu$,
namely $\psi(g)=\psi(\nu)$. Therefore, the inner product above
takes the form \be
\langle\psi,\phi\rangle=\frac{2}{\pi}\int\limits_0^{\pi} d\nu
\sin^2(\nu) \overline\psi(\nu)\phi(\nu)\,, \ee where we have
explicitly performed the $\theta$ and $\phi$ integration. In the
Lorentzian case the analog of the last step (in a formal
manipulation) leads to a trivial divergence due to the  noncompactness of the gauge group. It is clear that one can
regularize this divergence by simply dropping such integration. We
will  see this in detail in the following section.

Incidentally, notice that the characters can be expressed in terms
of linear combinations of powers of $\chi_{1/2}(g)$ which shows
that the most general gauge invariant functional of the
generalized connection is the observable (\ref{class-dirac-obs})
$O_2={\rm Tr}P
\exp(-\int_{S^1} \omega)$. The Dirac observable $O_1$ is quantized by
the self-adjoint operator $O_1=-\hbar^2 \Delta$, where $\Delta$
denotes the Laplacian on ${\rm SU}(2)$ , the characters $\chi_j(g)={\rm
Tr}[D^j(g)]$ are its eigenstates with eigenvalues $\hbar^2
j(j+1)$, namely \be O_1|j\rangle= \hbar^2 j(j+1) |j\rangle\,,\ee
where we have used Dirac bracket notation $\chi_j(g)\rightarrow
|j\rangle$. The Dirac observable $O_2$ acts by mutiplication, in
the spin network basis its action is given by \be
O_2|j\rangle=c_{j}^{\va(+)}
|j+{\vani\frac{1}{2}}\rangle+c_{j}^{\va(-)}
|j-{\vani\frac{1}{2}}\rangle\,,\ee where $c^{\va(\pm)}_j$ are the
corresponding Clebsh-Gordon coefficients. The gauge invariant
combination $O_3$ of $\phi^i$ and $g_{\Sigma}[\omega]$ can be
quantized using the commutator of $O_1$ and $O_2$ (Note that the
direct quantization of $O_3$ would require dealing with nontrivial
ordering issues).

\subsubsection{The Lorentzian case, $G={\rm SL}(2,\R)$}

 The Lorentzian case is more involved, and, to our knowledge,
has not been described in  the literature. The main technical
complication is the noncompactness of the gauge group which
implies that the standard nonperturbative techniques applicable
to standard gauge theories with compact groups need to be revised
 due to the appearance of divergences in the naive treatment.  The
main technical complication is due to the fact that, unlike the
compact case, class functions $\psi(g)=\psi(aga^{-1})$ are not
square integrable functions in the Haar measure of ${\rm SL}(2,\R)$. The
main difficulty with which one needs to deal is the definition of
the physical inner product for functions of ${\rm SL}(2,\R)/\sG$ so that
appropriate reality conditions are satisfied by the Dirac
observables. On the other hand the structure of the theory is
richer.

As we saw in the Riemannian case, physical states are
characterized by class functions which in turn can be thought  off
as functions of elements of the Cartan subgroups. The new feature
in the Lorentzian sector is that there are two inequivalent (under
conjugation) Cartan subgroups in ${\rm SL}(2,\R)$. On the one hand one
has $H_1={\rm U}(1)\in {\rm SL}(2,\R)$, given by rotations fixing an internal
time axis, namely
\be
H_1=\left\{g_{\nu}=\left[\begin{array}{ccc}\cos(\nu)&\sin(\nu)\\
-\sin(\nu)& \cos(\nu)\end{array}\right]\right\}\,, \label{h1}\ee
and $H_2\in {\rm SL}(2,\R)$ given by the subgroup of boosts fixing
some spacelike internal direction, explicitly
\be H_2=\left\{g_{\eta}=\left[\begin{array}{ccc} \exp(\eta)& 0\\
0&\exp(-\eta)\end{array}\right]\right\}\,. \label{h2}\ee This
means that conjugation $g\rightarrow aga^{-1}$ for fixed $g\in
{\rm SL}(2,\R)$ and arbitrary $a\in {\rm SL}(2,\R)$ generate
orbits (gauge orbits) which are labelled by elements of $H_1$,
$H_2$,
 respectively\footnote{More precisely, this is true for the so called regular
elements of ${\rm SL}(2,\R)$ which are an open subset of the full group
with complement of measure zero with respect to the Haar
measure~\cite{ruhl}.}.

Therefore, gauge invariant states $\Psi(g)=\Psi(aga^{-1})$ can be
characterized by two functions
\[\Psi[g]=\left\{\begin{array}{ccc} \psi_1(\nu)\ \ \
\mbox{for $[g]\in H_1$} \\ \psi_2(\eta)\ \ \
\mbox{for $[g]\in H_2$}\end{array}\right. \,.\]
As mentioned above the noncompactness of the gauge group implies
that gauge invariant states $\Psi(g)=\Psi(aga^{-1})$ are not
square integrable with respect to the Haar measure.  The reason is
that physical states are constant on noncompact adjoint orbits.
However, an inner product can be introduced in such a way that
these states are normalizable and the appropriate self-adjoint
property of observables holds.

This is easily done by mimicking what we did in the previous
section when finding an explicit parametrization of the Haar
measure in terms of the Cartan subgroup. It is still true that one
can write a regular element of ${\rm SL}(2,\R)$ as Abelian ``rotations
around an axis $\hat n$''. The main difference is that the unit
vector $\hat n$ can be either time like or space like. In other
words the (rotationally invariant) $2$-sphere of directions is now
replaced by the (${\rm SO}(2,1)$    invariant)
time like Hyperboloid
$h_{0}$, (given by the points in Minkowski internal spacetime
$x^{i}x^j\eta_{ij}=1$) together with the future and past space
like Hyperboloid $h_{\pm}$ (given by the points in Minkowski
internal spacetime\footnote{ Minkovski metric
is $\eta_{ij}={\rm diag}(-1,1,1)$.}
 $x^{i}x^j\eta_{ij}=-1$). The second difference
is that ``rotations around $\hat n$'' are now standard ${\rm U}(1)$
rotations (elements of $H_1$) only when $\hat n\in h_{\pm}$, wile
they are replaced by boosts (elements of $H_2$) when $\hat n\in
h_{0}$.

 Elements of $g\in {\rm SL}(2,\R)$ can be modelled by points on
three dimensional DeSitter spacetime (thought of as embedded in a
four dimensional flat spacetime of signature $(++--)$) as follows
\[g=\left(\begin{array}{ccc}x^0+x^3, x^1+ x^2\\ x^1-x^2, x^0-x^3\end{array}\right)
\ \ \ (x^0)^2-(x^1)^2+(x^2)^2-(x^3)^2=1\,.\]
In terms of these coordinates the invariant measure takes the
simple form
\[dg=dx^0 dx^1 dx^2 dx^3\ \delta((x^0)^2-(x^1)^2+(x^2)^2-(x^3)^2-1)\,.\]
Elements $g\in {\rm SL}(2,\R)$ equivalent by conjugation to elements in
$H_1\subset {\rm SL}(2,\R)$ are characterized by $|{\rm Tr}[g]|\le 2$.
It is easy to see that these elements can be described in terms of
hyperbolic coordinates as \ba && \nonumber x^0 = \cos[\nu]\\ &&
\nonumber x^1 = \sin[\nu]\sinh[\rho]\cos[\phi]\\&& \nonumber x^2 =
  \sin[\nu]\cosh[\rho]\\ && \nonumber x^3 = \sin[\nu]\sinh[\rho]\sin[\phi]\,,\ea
where $\rho\in \R^+$ and $\phi\in[0,2\pi]$ label points on
$h_{\pm}$. For these regular elements the invariant measure
becomes \be dg=\sinh(\rho)\sin^2(\nu)d\nu d\rho d\phi\,, \ \ \
{\rm for}\ \ \ [g]\in H_1\,. \label{uno}\ee Elements  $g\in
{\rm SL}(2,\R)$ equivalent by conjugation to elements in $H_2\subset
{\rm SL}(2,\R)$ are characterized by $|{\rm Tr}[g]|\ge 2$. In terms of
hyperbolic coordinates they are characterized by \ba && \nonumber
x^0 = \cosh[\eta]\\ && \nonumber  x^1 =
\sinh[\eta]\sin[\theta]\cosh[\rho]\\ && \nonumber x^2 =
\sinh[\eta]\sin[\theta]\sinh[\rho]\\ && \nonumber x^3 =
  \sinh[\eta]\cos[\theta], \ea where $\rho, \in \R^+$ and
$\theta \in[0,\pi]$ label points on $h_{0}$. Explicitly we have
${\rm Tr}[g]=2\cosh[\eta]$. The invariant measure becomes
\be
dg= \sin(\theta)\sinh^2(\eta)d\eta d\rho d\phi, \ \ \ {\rm for} \
\ \ [g]\in H_2\,.\label{dos}
\ee
Indeed the physical Hilbert space $\sH_{phys}=\sH_1\oplus\sH_2$  is
such that $\Psi(g)=(\psi_1(\nu),\psi_2(\eta))\in \sH_{phys}$ with
$\psi_1(\nu)\in \sH_1$ and $\psi_2(\eta)\in \sH_2$. The inner
product is respectively:
\[
\langle\psi,\phi\rangle_1=\int\limits_0^{2\pi} d\nu \sin^2(\nu) \overline\psi(\nu)\phi(\nu), \ \ \ {\rm and}\ \ \
\langle\psi,\phi\rangle_2=\int\limits_0^{\infty} d\eta
\sinh^2(\eta) \overline\psi(\eta)\phi(\eta)\,,
\]
 where the
integration measure is defined by dropping the redundant
integrations from the invariant measures (\ref{uno}) and
(\ref{dos}) respectively. The two Hilbert spaces $\sH_1$ and
$\sH_2$ are super-selection sectors. Harmonic analysis on
${\rm SL}(2,\R)$ implies that any function $f(g)\in {\sL^2({\rm SL}(2,\R))}$
can be written as:
\be
f(g)=\sum_{n\ge1} (2n-1) {\rm Tr}(f_n^+ D_n^+[g])+\sum_{n\ge 1}
(2n-1) {\rm Tr}(f_n^- D_n^-[g])+\int_0^\infty ds \mu(s) {\rm
Tr}(f_s D_s[g]) \,.
\ee
As in the Euclidean case, physical states can
be spanned in terms of characters. From the point of view of the
space $\sL^2({\rm SL}(2,\R))$ these states are distributional. The
restriction of the characters to $H_1$ and $H_2$ give an
orthonormal basis of $\sH_1$ and $\sH_2$ respectively of
eigen-states of the Dirac observable
$O_1=\phi\cdot\phi=-\hbar^2\Delta$. In $\sH_1$ the Laplacian takes
the explicit form
$\Delta=\sin(\nu)^{-1}\partial_{\nu}^2\sin(\nu)+1/4$ and the
characters are $\chi^{\pm}_n(\nu)=\pm \exp(\pm
i(n-1)\nu)/(2i\sin(\nu))$ and the eigenvalues of $O_1$ are given
by $\hbar^2 (n-{\vani\frac{1}{2}})(n-{\vani \frac{3}{2}})$.
Therefore, the spin network states $\chi^{\pm}_n(\nu)\rightarrow
|n,\pm\rangle$, which satisfy
\[\langle n,+|m,+\rangle_1=\langle n,-|m,-\rangle_1=\delta_{nm}\ \ \  {\rm
and}\ \ \ \langle n,+|m,-\rangle_1=0\,,\] are a basis for $\sH_1$
diagonalizing $O_1$
\be
O_1|n\rangle= -\hbar^2
(n-{\vani\frac{1}{2}})(n-{\vani
\frac{3}{2}})|n\rangle\,.\label{gapd}
\ee
The Dirac observable $O_2$
acts by mutiplication by $O_2=2\cos(\nu)$ in $\sH_1$. In the spin
network basis its action is given by \be O_2|n,\pm \rangle= |n\pm
1,\pm\rangle+|n\mp 1,\pm\rangle\,.\ee

In $\sH_2$ the Laplacian takes the explicit form
$\Delta=-\sinh(\eta)^{-1}\partial_{\eta}^2\sinh(\eta)+1/4$ and the
characters are $\chi_s(\eta)=\pm \cos(s\eta)/|\sinh(\eta)|$ and
the eigenvalues of $O_1$ are given by $\hbar^2 (s^2+{\vani
\frac{1}{4}})$. Now spin network states are labelled by a real
parameter $\chi_s(\nu)\rightarrow |s\rangle$. We have \be
O_1|s\rangle= \hbar^2 (s^2+{\vani
\frac{1}{4}})|s\rangle\,,\label{gap}\ee The Dirac observable
$O_2=\cosh(\eta)$ acts by multiplication. Its  action on
spin network states is not a spin network state.

\subsection{Time-gauge quantization}\label{Time-gauge quantization}

The purpose of~\cite{CLMPR} and~\cite{paper-prepa} is a
quantization along the lines of loop quantization in 1+3 dimensions with a
time-gauge fixing, in the presumably simpler case of the (1+1)-dimensional
Jackiw-Teitelboim model. However, the procedure still remains somewhat more
complicated than in the other cases studied in the present paper. The
construction of the kinematical Hilbert space is based on the wave
functionals defined on the configuration space spanned by the
"holonomies" of
the scalar fields $\phi^1$ and $\phi^2$ defined in Subsection
\ref{Time-gauge reduction}, the ``polymer-like" scalar product
used there leading to nonseparability of the Hilbert space~\cite{scalar-field}.
The conjugate fields $\omega^1$ and $\omega^2$ are represented as
functional  differential operators, which are diagonal in a spin-network-
like orthonormal basis.

The construction of operators representing the classical constraints
(\ref{g1}) goes through a cell regularization, and the hope is to check
the algebra (\ref{alg3}) at the quantum level, the final task being that of
solving the quantum constraints.


\subsection{Quantization in the $\phi\cdot n=0$
gauge.}\label{phi-dot-n-quant}

Recall that the $\phi\cdot n=0$ gauge in the Lorentzian case, as described
classically in
Section (\ref{Partial gauge fixings}), lets us with two first class
constraints $C_0$ and $C_1$ (\ref{C-first-class}) in the $\phi\cdot\phi>0$
case,
$C_0$ generating U(1) internal rotations in the $(\phi^1,\,\phi^2)$ plane,
and $C_1$ a boost leaving the vector $\phi=(0,\phi^1,\phi^2)$ invariant. In the case
$\phi\cdot\phi<0$, we are left with one constraint $g_0$ (\ref{g-0-constraint})
generating  U(1) internal rotations leaving the vector $\phi=(\phi,0,0)$
invariant.
In the former case, to the transformation generated by $C_1$ is associated the
Abelian connection $Q$ (\ref{abelian-conn}).
In the latter case, to the U(1) invariance is
associated the connection $\omega=\omega^0$.
In the spirit of loop quantization, one must thus
choose the holonomies $H_I$ of $Q$ or $\omega$ along intervalls $I$ of $S^1$
as the configuration space variables
(from the
connections we go to the generalized connections).
These Abelian holonomies  are given by
\be
H_I=\exp(-\int_I dx {\cal T}(x) \taus)\,,
\label{holonomy-I}\ee
with
\[\begin{array}{lll}
\cal T=Q\,,\quad &\taus= \left[\begin{array}{ccc}-1&0\\
0& 1\end{array}\right]\quad &\mbox{for the sector }
\phi\cdot\phi>0\,,\\[3mm]
\cal T=\omega\,,\quad &\taus = \left[\begin{array}{ccc}0&-1\\
1& 0\end{array}\right]\quad &\mbox{for the sector }\phi\cdot\phi<0\,.
\end{array}\]
Notice that these are the generators whose exponentiation leads to
group elements of the form \ref{h2} (a boost) and \ref{h1}
(a rotation), respectively.
The holonomy $H=H_{S^1}$ is obviously a Dirac observable, element
of the adjoint representation of the
group generated by $C_1$ or $g_0$.

Let us know discuss the two sectors separately.

\subsubsection{ Sector $\phi\cdot\phi>0$}

It will be convenient to use the invariant (under
the action of $C_0$) variables $(\ref{u1variables})$, namely
\[
\Pi\equiv\frac{1}{2}\phi^A\phi^A \ \ \ {\rm and} \ \ \
Q\equiv\frac{\phi^A\omega^A}{\phi^C\phi^C}\,.
\]
The advantage of
these variables is that $\Pi$ is nothing else but $O_1$
(\ref{class-dirac-obs}), one of the Dirac observables, and $Q$
is the Abelian connection discussed above,
transforming  under the remaining Abelian gauge
symmetry generated by the constraint $C_1$  (boosts) as:
\[
\{Q,C_1(b)\}_{\va D} = -\partial b\,.
\]
(See (\ref{C-first-class}, \ref{C(a)})).
A finite transformation of $H_I$ as given by (\ref{holonomy-I}) reads:
\[
 H_I' = \exp (\ts(b_t-b_s)) H_I\,,
\]
where $b_s$ and $b_t$ are the values of $b$ at the ends
(source/target) of the interval $I$. Therefore, the holonomy
around the space, $H:=H_{S^1}$, being gauge invariant is a Dirac observable.
It takes the form
\be
H= \exp(-\taus\eta) =
\left[\begin{array}{ccc}\exp(\eta)& 0\\
0& \exp(-\eta) \end{array}\right]\,,
\quad\mbox{with} \ \eta=\int_{S^1}dx Q(x)\ \mbox{and}\
\tau_{\va \! \star}\equiv \left[\begin{array}{ccc}-1&0\\
0& 1\end{array}\right]\,.
\label{H(eta)}\ee
From the classical Dirac bracket algebra (\ref{Dirac-bracket})
follows
\be
\{\Pi(x),H\}_{\va D} = \taus H\,,
\label{qqq}\ee
and consequently
\be
\{\Pi, \eta\}_{\va D} = -1\,.
\label{pi-eta-bracket}\ee
This suggests to take $\Pi$ and $\eta$ as the classical phase space coordinates,
with $\Pi\in\R_+$ (real positive) and $\eta\in\R$. (Owing to the
constancy of $\Pi$ following from the
constraint $C_1$, we can take for $\Pi$ the value of $\Pi(x)$ at any point
$x\in C_1$.)

We can quantize this sector in the
conventional Schroedinger scheme. Here, elements of the physical Hilbert
space are functions $\Phi(\eta)$ -- with $\eta$ defined in
(\ref{H(eta)}) -- belonging to ${\rm L}_2(\R,d\eta)$, and $\Pi$ is
represented by the operator $\hat\Pi = -i\hbar d/d\eta$. Eigenfunctions
of $\hat\Pi$ are unnormalizable ``plane waves'' $\exp(i\rho\eta/\hbar)$. Restricting to
``wave packets'' of positive frequency $\rho>0$ we recover the set of
positive real number as the (now continuous) spectrum of $\hat\Pi$.
The spectrum of $\Pi=O_1$ is thus constituted by the positive real numbers in agreement
with the results of the previous section, but it does not quite
coincide with it for small values of $s$,
as there is no discrete gap here (in contrast
with (\ref{gap})). However there is agreement
in the asymptotic regime.

Alternatively, as $H$ is an element of the Abelian boost group, parametrized by the real
number $\eta$ one could explore the quantization based on the so-called polymer
representations~\cite{polymersAsh,polymersCorichi}
Our task is to construct a physical Hilbert space $\Hp$ as a
representation of the quantum algebra $[\hat\Pi,\hat\eta]= -i\hbar$ or,
better:
\be
[\hat\Pi,\hat h]= \hbar\hat h\,,\quad\mbox{where}\ h = \exp(i\eta)\,,
\label{q-alg-P-eta}
\ee
corresponding to the classical algebra (\ref{pi-eta-bracket}).
The elements of $\Hp$ will be taken as functions of the boost group,
which we parametrize by real numbers $\rho$
considered as elements of the Bohr compactification~\cite{Rudin} $\RB$ of
$\R$. Accordingly, the integration measure for the ``almost periodic
functions''
\[
\Phi_s(\rho)=\exp(-\frac{i}{\hbar}s\rho)\,,\quad s\in\R\,,
\]
 is given by
\be
\int_\RB d\mu(\rho) \Phi_s(\rho)=\delta_{s0}\,.
\label{Bohr-measure}\ee
The ``cylindrical vector space'' $\sH_{\rm cyl}$ is defined as the set of all
finite linear combinations of almost periodic functions, and a
hermitean scalar product is defined with the
help of the integration measure (\ref{Bohr-measure}). Hence, in
particular:
\be
\vev{\Phi_s|\Phi_{t}} = \delta_{st}\,.
\label{cyl-scalar-product}\ee
The action of the operator  $\hat\eta$ is defined through the action of
its almost periodic counterparts
$\Phi_t(\rho)$ for any $t\in\R$:
\[
\hat\Phi_t\Phi_s(\rho) = \Phi_{t+s}(\rho)\,,
\]
whereas that of $\hat\Pi$ is  defined by
\[
\hat\Pi \Phi_s(\rho) = -i\hbar \frac{d}{d\rho} \Phi(\rho)\,,
\]
in agreement with the algebra (\ref{q-alg-P-eta}).

We observe that the $\Phi_s$ are eigenvectors of $\hat\Pi$ with
eigenvalue $s$. Since this operator owes to be positive beyond of being
self-adjoint, we define $\sH_{{\rm cyl}\,+}$ as the space generated by
the restricted basis $\{\Phi_s\,;\ s\ge0\}$ -- in analogy with the
separation of the positive and negative frequency parts in the relativistic
quantum theory of free fields.
Finally, the
physical Hilbert space $\Hp$ is defined as the Cauchy completion of
$\sH_{{\rm cyl}\,+}$ with respect to the norm
induced by the scalar product (\ref{cyl-scalar-product}).
Possessing an uncountable orthonormal basis, $\Hp$ is nonseparable.
We note that the spectrum of $\hat\Pi$ coincide with the one found in the
Schroedinger representation, yet it
must be considered as "discrete" as the corresponding
eigenvectors have finite norm.

\subsubsection{ Sector $\phi\cdot\phi<0$}

In this sector, as discussed in Subsection (\ref{time-like sector}),
the classical phase space variables are
$\omega=\omega^0$ and $\phi=\phi^0$, obeying the canonical Dirac bracket
algebra (\ref{temp-case-Dirac-bracket}). The constraint $g_0$ (\ref{g-0-constraint})
lets $\phi$ to be a constant,
and $\omega$ transforms under $g_0$ as the
connection associated to the residual group U(1) of gauge
transformations preserving the gauge fixing conditions $\phi^1=\phi^2=0$. Thus
the classical Dirac observables are $\phi$ -- taken at an arbitrary valor
of the space coordinate $x$ -- and the holonomy along the space slice:
\[
H=\exp(\taus\xi) = \left[\begin{array}{ccc}\cos(\xi)& \sin(\xi)\\
-\sin(\xi)& \cos(\xi)\end{array}\right]
\,,\quad \xi=\int_{S^1}dx\om(x)\,,\quad
\taus=\left[\begin{array}{ccc}0&-1\\
1& 0\end{array}\right]\,.
\]
It is convenient to perform the quantization in terms of the variables $\phi$ and
$\xi$. From now on we switch
emphasis from $H(\xi)\rightarrow h(\xi):= \exp (i\xi)$.
From the classical Dirac bracket
$\{\phi(x),h\}_{\va D} = ih$, we define the corresponding quantum commutator
as
\[
[\hat\phi,\hat h]=\hbar\hat h\,.
\]
With this choice of variables, elements of the physical
Hilbert space are continuous functions of U(1), i.e.,
$\Phi(\theta)$, with $\theta\in[0,2\pi]$, and $h(\theta)$ acts simply
by multiplication.  There is a spin network basis given
by the unitary irreducible representations of U(1), explicitly:
\[
\Phi_n(\theta)=\frac{1}{\sqrt{2\pi}} \exp(i n
\theta)\,,\quad \mbox{for all\ }n\in \Z  \,.
\]
The spectrum of $\hat\phi$ is discrete, given by its eigenvalues $n\hbar$.
Thus the spectrum of $O_1=\hat\phi\cdot\hat\phi$
is in agreement with the
results of the previous section but has eigenvalues that differ
from those corresponding to (\ref{gapd})) for small value of $n$.
However, the
eigenvalues of $O_1$ approach each other in the large (in Planck
units) eigenvalue limit.


\subsection{Quantization after totally reducing}
The  Dirac bracket algebra given by Eq. (\ref{XP})
tells us that the  reduced phase space of the
  Jackiw-Teitelboim model corresponds to that of a system with a
single degree of freedom. There are however prequantization
conditions that are expected to make the spectra of quantum
observables compatible (at least in the large eigenvalue limit) with
the results of the previous sections (for detail see
\cite{Strobl:1999wv, martin}).

\section{ Conclusions}

This paper is  divided in two parts. In the first part we performed the
canonical analysis of the JT model and studied the phase space structure of
the theory. We showed that there are dynamically independent sectors
corresponding to the cases $\phi\cdot\phi>0$ (space-like), $\phi\cdot\phi=0$ (null) and
$\phi\cdot\phi>0$ (time-like). The system has no local degrees of freedom. However,
for $M=S^1\times \R$ there is one global topological degree of freedom in the
time-like and space-like sectors respectively. The null sector is
special, classical (physically inequivalent) solutions are labeled by a
discrete parameter (winding number) \cite{martin}.

A partial gauge fixing allows, in the space-like case, to reduce the DeSitter
gauge symmetry of the JT model to the two dimensional  diffeomorphism
invariance of gravity. We explicitly showed in that case how, after partial
gauge fixing, the remaining
first class constraints relate to the generators of two  Virasoro symmetries,
and the familiar diffeomorphism and scalar constraints of gravitational
theories.

In the second part we studied the quantization of the JT model using
background independent techniques. We first performed the quantization of the
model without the introduction of gauge fixing. Even though this was well known in the
Riemannian case, the Lorentzian case presented some technical difficulties
related to the  noncompactness of the gauge group. The difficulties were
overcome using group averaging techniques which  naturally lead to two possible Hilbert space
representation of the fundamental observables. These inequivalent quantum
theories have both physical interpretation as they are in one-to-one
correspondence with the superselection sectors  $\phi\cdot\phi>0$
(space-like), and $\phi\cdot\phi>0$ (time-like).

Alternatively, one can partially (or totally)
reduce the gauge freedom of the system at the classical level by
introducing gauge conditions, and explore afterwards  the
quantization of the reduced system. We have explicitly shown that
representations of the basic fields obtained  following this alternative avenue
are not equivalent to the the previous ones.  The main results of
 the sencond
part  are summarized in Table
(\ref{table}),
which shows the various inequivalent sectors of the quantum
Jackiw-Teitelboim theory.
\begin{table}[hbt]
\begin{tabular}{|l||l|l||l|l|}
\hline
 &\multicolumn{2}{l|}{Riemannian}&\multicolumn{2}{l|}{Lorentzian}\\
\cline{2-5}
 &Spect. $O_1$& &Spect. $O_1$ in Sector I
     &Spect. $O_1$ in  Sector II\\
\hline\hline No gauge-fixing& $\hbar^2 n(n+1)$ for $n\in\Z/2$&
&$-\hbar^2 (n-{\vani\frac{1}{2}})(n-{\vani \frac{3}{2}})$ for
$n\in\Z$&$\hbar^2
(s^2+{\vani \frac{1}{4}})$ for $s\in\R^+$\\
$n\cdot\phi=0$ gauge&  $\hbar^2 n^2$ for $n\in\Z$&&  $-\hbar^2 n^2$ for $n\in\Z$& $\hbar^2
s^2$ for $s\in\R^+$ \\
Time gauge & ? & &?&?\\
\hline
\end{tabular}
\caption{Spectrum of the observable $O_1$ defined in Eq. (\ref{class-dirac-obs}),
according to various quantization schemes.}
\label{table}\end{table}
We have called I and II the sectors (phases) of the Lorentzian theory which are
described by the Hilbert spaces $\sH_1$ and $\sH_2$ in the no gauge
fixing part (Section \ref{Dirac_quant}), and by the $\phi\cdot\phi<0$
and $\phi\cdot\phi>0$ sectors in the $n\cdot\phi=0$ gauge (Section
(\ref{phi-dot-n-quant}).

We observe that the spectrum of $O_1$ in the
nongauge-fixed and $n\cdot\phi$-gauge coincide in the large
eigenvalue limit in both the Riemannian and the Lorentzian phase I.
This is compatible with a
common  semi-classical limit. But, in the Lorentzian phase II
 the spectrum is that of all positive real numbers,
``discrete'' or ``continous'' depending on the quantization being ``polymeric'' or
of the ``Schroedinger'' type.
Finally, the spectrum of $O_1$ is completely changed (continuous) in
the fully reduced case, for the Riemannian and for both phases of the
 Lorentzian
theory. The gauge group structure vanishing in
this case implies that the kind of representations used in
the former cases are not even available.
As discussed in Section \ref{Time-gauge quantization}, one could use a polymer
like representation to recover a ``discrete'' spectrum but the
microscopic Planckian structure is lost in the fully reduced
setting.

 We have not attempted the quantization of the null sector. This problem
seems quite subtle. We notice that the group averaging quantization of
Section \ref{Dirac_quant} seems to miss that sector. It would be desirable to
fully understand the quantum nature of the null sector. At this stage this
is beyond the scope of this paper.

\section{Note added on the quantization of the null sector (not in
published paper)}

In the case of the null sector one can first gauge fix 
$\phi^0=1$. This reduces the cone $\phi\cdot\phi=0$ to a circle. The
irreducible first class constraints are $C_0=\epsilon_{AB} \omega^A
\phi^B$ and $g_4=\phi^A\phi_A -1$, and the index $A,B=1,2$. The
constraint $C_0$ generates $U(1)$ rotations. There are no local
degrees of freedom. There is a $U(1)$ connection $\Gamma$---the
analog of the spin connection---defined in equation III.22,
explicitly \be \Gamma=\frac{\epsilon_{AB} \phi^A
d\phi^B}{\phi_C\phi^C}\ee This connection is very easy to picture
geometrically. If you use polar field coordinates and define \be
r^2=\phi^A\phi_A\ee and \be \phi^1=r \cos(\theta)\ \ \  \phi^2=r
\sin(\theta) \ee  then the constraints simply become
$C_0=p_{\theta}$ and the other one $g_4=r-1$ and the $U(1)$
connection defined in III.22 is simply \be \Gamma=d\theta. \ee
Physical configurations are maps from $S^1$ to $S^1$ sending $x\in
\Sigma \approx S^1$ (space slice) into $\theta(x)$. One important
Dirac observable is the one that measures the winding of this map.
It is simply given by \be W=\frac{1}{2\pi} \int_{\Sigma} \Gamma \ee
i.e. the winding number of the map. Now we come to the important
point: W commutes with all observables in the theory. Therefore W is
not quantized. The quantum theory is a tower of one dimensional
Hilbert spaces $\sH_W$ labeled by the winding number W that remains
classical!


\vspace{4mm}

\noindent  {\bf Acknowledgements.} One of the authors (A.P.) thanks
the Coordena\c c\~ao de Aperfei\c coamento de Pessoal de N{\'\i}vel
Superior (CAPES, Brazil) for financial support under the PVE program
and the Physics Department of the Universidade Federal do
Esp{\'\i}rito Santo for hospitality. We thank M. Bojowald for
clarifying remarks and for pointing out a confusion in the counting
of degrees of freedom of the null sector in a early version of the
manuscript.



\end{document}